\documentclass[12pt]{iopart}
\usepackage{iopams}  
\usepackage{bm}
\usepackage{bbm}
\usepackage{graphics}
\usepackage{cite,hyperref}

\begin{document}

\title[\small{Microscopic approach to field dissipation in the Jaynes-Cummings model}]{Microscopic approach to field dissipation in the Jaynes-Cummings model}

\author{C. A. Gonz\'alez-Guti\'errez}
\address{Instituto de Ciencias F\'isicas, 
	Universidad Nacional Aut\'onoma de M\'exico, Avenida Universidad s/n, 62210 Cuernavaca, Morelos, M\'exico}
\ead{carlosag@fis.unam.mx, carlosgg04@gmail.com}

\author{D. Sol\'is-Valles}
\address{Tecnologico de Monterrey, Escuela de Ingenier\'ia y Ciencias, Ave. Eugenio Garza Sada 2501, Monterrey, N.L., M\'exico, 64849.}
\ead{A00367476@itesm.mx}

\author{B. M. Rodr\'iguez-Lara}
\address{Tecnologico de Monterrey, Escuela de Ingenier\'ia y Ciencias, Ave. Eugenio Garza Sada 2501, Monterrey, N.L., M\'exico, 64849.  \\
Instituto Nacional de Astrof\'isica, \'Optica y Electr\'onica, Calle Luis Enrique Erro No. 1, Sta. Ma. Tonantzintla, Pue. CP 72840, M\'exico}
\ead{bmlara@itesm.mx, bmlara@inaoep.mx}

\vspace{10pt}

\begin{abstract}
We use the microscopic derivation of the Jaynes-Cummings model master equation under field losses to study the dynamics of initial field states beyond the single-excitation manifold. 
We show that field-qubit detuning, as well as finite temperature, modify the effective decay rate in the model using entropy measures, like qubit-field purity and von Neumann entropy of the field, for initial Fock states. For initial semi-classical states of the field, we show that the microscopic approach, in phase space, provides an evolution to thermal equilibrium that is smoother than the one provided by the standard phenomenological approach.
\end{abstract}

\pacs{42.50.Ar, 03.67.Lx, 42.50.Pq, 85.25.-j}
%
%
%
\maketitle
%
%

\section{Introduction}

The Jaynes-Cummings (JC) model \cite{Jaynes1963p89}, describing the interaction of a single bosonic mode with a two-level system, plays a key role in our understanding of interaction between radiation and matter.
It is of central importance for the description of quantum effects, for example, the existence of Rabi oscillations for Fock field states \cite{Jaynes1963p89} and the collapse and revival of the atomic inversion in the presence of coherent fields \cite{Eberly1980p1323}, and constitutes a basic building block for the implementation of quantum gates \cite{Nielsen2000}. 
The model has been implemented in a variety of experimental platforms \cite{Raimond2001p565,Leibfried2003p281,Xiang2013p623}, where the unavoidable effect of the environment over  closed-system dynamics is observed as a deterioration, or even complete suppression, of the expected quantum phenomena \cite{Rempe1987p353,Cirac1994p1202,Meekhof1996p1796,Brune1996p1800}.
Thus, an adequate description of loss-mechanisms in different physical scenarios became essential to compare with experimental results, and lead to the proposal and study of different decoherence and dissipation models in the literature \cite{Puri1986p3610,Eiselt1989p351,Eiselt1991p346,Barnett2007p2033,Chiorescu2004p159,Wallraff2004p162,Wilczewski2009p033836,Wilczewski2009p013802}.

Here, we are interested in the microscopic approach to field dissipation in the standard Jaynes-Cummings model. 
The microscopic approach has demonstrated fundamental dynamical differences with the usual phenomenological approach for the single excitation manifold of the Jaynes-Cummings at zero \cite{Scala2007p013811} and finite \cite{Scala2007p14527} temperature.
Both, the microscopic and phenomenological models of dissipation make use of the Born-Markov approximation, that considers a memory-less environment that couples weakly to the system. They differ on the fact that the microscopic approach uses the dressed state basis that diagonalizes the JC model in order to derive the effective master equation, while the phenomenological approach uses the microscopic master equation derived for just a dissipative field mode.

In the following, we review the microscopic derivation of the master equation for field dissipation in the Jaynes-Cummings model, and provide analytical expressions for the state evolution of the system that agree with previous results.
Then, we compare the dynamics under this master equation and the standard phenomenological approach beyond the single excitation manifold at zero and finite temperature with a flat environment. 
In particular, we demonstrate the time evolution due to initial number and coherent field states through standard observables, like atomic inversion, mean photon number, entropy-related measures, such as purity and von Neumann entropy, and phase space quantities, like quadratures of the field and Husimi Q-function.
Finally, we provide our conclusions and perspective for the description of dissipation for radiation matter interaction in strong-coupling regimes \cite{Boite2016p033827}.

\section{Microscopic approach for the open JC model}

\begin{figure}
	\centering
	\includegraphics{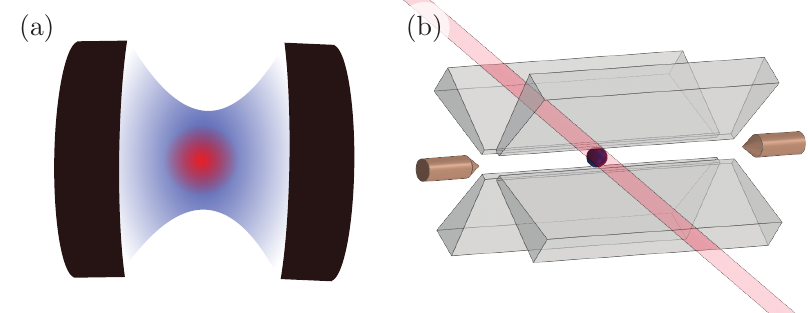}	
	\caption{Schematics for two experimental realizations of the Jaynes-Cummings model, (a) cavity-QED, and (b) ion-trap-QED.} \label{fig:Fig1}
\end{figure}

We follow the formal microscopic derivation of the Markovian master equation for the JC model \cite{Scala2007p013811}, and start from the standard JC Hamiltonian \cite{Jaynes1963p89},
\begin{equation}\label{JCModel}
\hat H_{JC}=\frac{\omega_{0}}{2} \hat\sigma_{z} + \omega \hat{a}^{\dagger} \hat{a}   +g\left(\hat a\hat \sigma_{+}+\hat a^{\dagger}\hat \sigma_{-}\right),
\end{equation}
describing a two-level system, a qubit, with transition frequency $\omega_{0}$ and modeled by the standard atomic inversion operator, $\hat{\sigma}_{z}$, and lowering (raising), $\hat{\sigma}_{-}$ ($\hat{\sigma}_{+}$), operators, interacting with a boson field with frequency $\omega$, described by the annihilation (creation) operator $\hat{a}$ ($\hat{a}^\dagger$); the strength of the interaction is provided by the coupling parameter $g$.
The JC model assumes near resonance, $\omega \sim \omega_{0}$, and weak coupling, $g \ll \omega, \omega_{0}$. 
It relates to experimental realizations in cavity-QED \cite{Raimond2001p565}, trapped-ion-QED \cite{Leibfried2003p281}, circuit-QED and more \cite{Xiang2013p623}, Fig \ref{fig:Fig1}.

The JC model is a typical example of an  integrable system, it conserves the total number of excitations $\hat{N}= \hat{a}^{\dagger} \hat{a} + \left( 1 + \hat{\sigma}_{2} \right) /2$. It has a ground state provided by the boson vacuum and the qubit ground state, 
\begin{eqnarray}
\vert \epsilon_{0} \rangle = \vert 0, g \rangle,
\end{eqnarray}
with zero total excitation number, $\langle \hat{N} \rangle = 0 $, and the rest of eigenstates are given by the dressed state basis \cite{Gerry2005},
\begin{eqnarray}
\vert \epsilon_{n,+} \rangle &=& c_{n} \vert n,e \rangle + s_{n} \vert n+1,g \rangle,\nonumber  \\
\vert \epsilon_{n,-} \rangle &=&-s_{n} \vert n,e \rangle + c_{n} \vert n+1,g \rangle, \label{dressed}
\end{eqnarray}
which define subspaces with mean total excitation $\langle \hat{N} \rangle = n + 1 $, for integer index $n=0,1,2,\ldots$. The normalization coefficients are given by $c_{n} = \cos \left( \theta_{n} /2 \right)$, $s_{n} = \sin \left( \theta_{n} /2 \right)$, and the rotation angle $\theta_{n} = \arctan 2 g \sqrt{n+1} / \Delta$, where the detuning is defined by $\Delta = \omega_{0} - \omega$.
The energy spectrum,
\begin{eqnarray}
\epsilon_{0} &=& -  \frac{\omega_{0}}{2}, \nonumber \\
\epsilon_{n,\pm} &=& \left( n + \frac{1}{2} \right) \omega \pm \frac{\Omega_{n}}{2} ,
\end{eqnarray}
is given in terms of the Rabi frequency, $\Omega_{n} = \sqrt{ \Delta^{2} + 4 g^2 \left( n+1 \right)}$.

Now, we follow the standard formalism for open quantum systems \cite{Breuer2002}.
In other words, we model the environment as a collection of non-interacting bosons, $\hat H_{B} = \sum_{k} \omega_k^{\mathstrut } \hat{b}_{k}^{\dagger} \hat{b}_k^{\mathstrut }$, that bilinearly couple to the field via the interaction Hamiltonian, $\hat H_{I} = \hat{X}\hat  X_{B} $ with $\hat{X}= \hat{a}^{\dagger} + \hat{a}$ and $\hat{X}_{B}= \sum_{k} g_{k} \left(\hat{b}_{k}^{\dagger} + \hat{b}_{k} \right)$.
Then, we use the eigenmode decomposition, $\hat{X}(\nu) = \sum_{\nu} \hat{\Pi}(\epsilon) \hat{X}~ \hat{\Pi}(\epsilon^{\prime})$, in terms of  the projection operator $\hat{\Pi}(\epsilon)$ onto the dressed  subspace with effective frequency $\epsilon$ and  frequency difference $\nu= \epsilon^{\prime} - \epsilon$,
\begin{eqnarray}
\hat{X}(\nu) = \sum_{\epsilon^{\prime} - \epsilon = \nu} \langle \epsilon \vert \hat{X} \vert \epsilon^{\prime} \rangle ~ \vert \epsilon \rangle \langle \epsilon^{\prime} \vert.
\end{eqnarray}
This provides us with the explicit form of the boson field operator $\hat{X}$, in terms of the Bohr eigenfrequencies of the central system, such that the jump operators for the JC ladder become,
\begin{eqnarray}\label{jump::operators}
\hat{X}(\epsilon_{0,\pm}-\epsilon_{0}) &=& s_{0} \vert \epsilon_0 \rangle \langle \epsilon_{0,+} \vert  + c_{0} \vert \epsilon_0 \rangle \langle \epsilon_{0,-} \vert, \nonumber \\
\hat{X}(\epsilon_{n^{\prime},+}-\epsilon_{n,+}) &=& \delta_{n,n^{\prime}-1}
\left[c_{n} c_{n+1}\sqrt{n+1} + s_{n}s_{n+1}\sqrt{n+2}\right] 
\vert \epsilon_{n,+} \rangle \langle \epsilon_{n+1,+} \vert, \nonumber \\
\hat{X}(\epsilon_{n^{\prime},-}-\epsilon_{n,-}) &=& \delta_{n,n^{\prime}-1} \left[s_{n}s_{n+1}\sqrt{n+1}+ c_{n}c_{n+1}\sqrt{n+2}\right] \vert \epsilon_{n,-}\rangle \langle\epsilon_{n+1,-}\vert, \nonumber \\
\hat{X}(\epsilon_{n^{\prime},\pm}-\epsilon_{n,\mp}) &=& \delta_{n,n^{\prime}-1}
\left[s_{n}c_{n+1}\sqrt{n+2}-c_{n}s_{n+1}\sqrt{n+1}\right] 
\vert \epsilon_{n,\pm}\rangle \langle\epsilon_{n+1,\mp} \vert. \nonumber \\
\end{eqnarray}

Writing down the von Neumann equation for the the total density operator in the interaction picture
with the reference free Hamiltonian $\hat H_{0}=\hat H_{JC}+\hat H_{B}$, using the Born-Markov and rotating
wave approximations (RWA), and taking the average over the degrees of freedom of the environment
trough the partial trace operation, we can obtain the following master equation in the Schrödinger picture,
\begin{eqnarray}\label{general::master::eq}
\dot{\rho}(t)&=&-{i}[\hat H_{JC},\rho(t)]+\sum_{\nu>0}\gamma(\nu)
\left[ \hat{X}(\nu)\rho(t)\hat{X}^{\dagger}(\nu)-\frac{1}{2}\{\hat{X}^{\dagger}(\nu) \hat{X}(\nu),\rho(t)\}\right] \nonumber\\
&& + \sum_{\nu>0}\gamma(-\nu) \left[\hat{X}^{\dagger}(\nu)\rho(t)\hat{X}(\nu)-\frac{1}{2}\{\hat{X}(\nu) \hat{X}^{\dagger}(\nu),\rho(t)\}\right].
\end{eqnarray}
Note that the RWA is valid only for couplings larger than the decay rate, $2g\gg\gamma$. 
The effective frequency-dependent decay rates are given by the Fourier transform,
\begin{eqnarray}
\gamma(\nu)&=&\int _{0}^{\infty} ds ~ e^{i \nu s} ~ \mathrm{Tr}_{B} \left[  \hat{X}_{B}^{\dagger}(s) \hat{X}_{B}(0) \right], \nonumber \\
&=&\left\{
\begin{array}{ll}
\vert g(\nu) \vert^2 D(\nu) \left[  1 + \bar{n}\left( \nu \right)\right] , & \nu > 0 \\
\vert g(\vert \nu \vert ) \vert^2  D(\vert \nu \vert) \bar{n}\left( \vert \nu \vert \right) , & \nu < 0,
\end{array} \right.
\end{eqnarray}
with the continuum coupling distribution, $g(\nu)$, and the density of modes, $D(\nu)$, providing the environment spectral density, $\vert g(\nu) \vert^2 D(\nu)$; for example, a flat environment has a constant spectral density equal to the common decay rate, $\vert g(\nu) \vert^2 D(\nu) = \gamma$.  Finally, the average number of thermal bosons in the environment is defined by $\bar{n}(\nu) = 1 / \left( e^{\nu / k_{B} T} - 1 \right)$, with Boltzmann constant $k_{B}$ and finite temperature $T$.

\begin{figure}
	\centering
	\includegraphics{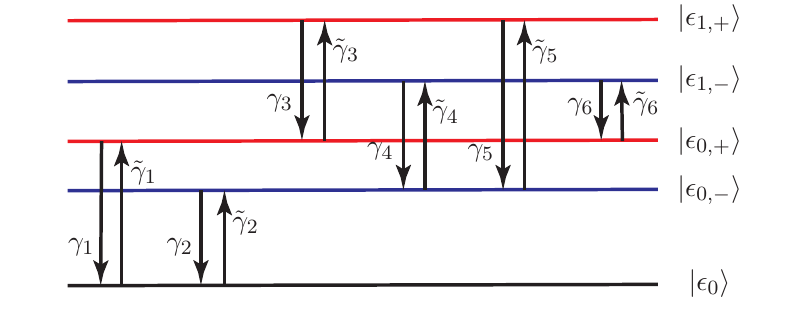}	
	\caption{A schematic example of some allowed transitions in the dressed energy ladder, decay and excitation channels, due to a finite temperature environment.} \label{fig:Fig2}
\end{figure}

In order to provide an explicit working form, we consider the microscopic master equation for the JC model interacting with a flat thermal bath at finite temperature, 
\begin{eqnarray}\label{eq:DSME}
\dot{\rho}(t)&=& - i [\hat H_{JC},\rho(t)] + \gamma_{1} s^{2}_{0} \hat{\mathcal{D}} (\vert \epsilon_{0} \rangle \langle \epsilon_{0,+}\vert )
+\gamma_{2}c^{2}_{0}\hat{\mathcal{D}} (\vert\epsilon_{0} \rangle \langle \epsilon_{0,-}\vert )
\nonumber \\
&& + \sum_{n=0}^{\infty}\gamma_{3}a_{n}^2 \hat{\mathcal{D}} (\vert\epsilon_{n,+} \rangle \langle \epsilon_{n+1,+}\vert )
+ \sum_{n=0}^{\infty}\gamma_{4}b_{n}^2 \hat{\mathcal{D}} (\vert\epsilon_{n,-} \rangle \langle \epsilon_{n+1,-}\vert ) \nonumber \\
&&+ \sum_{n=0}^{\infty}\gamma_{5}d_{n}^2 \hat{\mathcal{D}} (\vert\epsilon_{n,-} \rangle \langle \epsilon_{n+1,+}\vert ) 
+ \sum_{n=0}^{\infty}\gamma_{6}d_{n}^2 \hat{\mathcal{D}} (\vert\epsilon_{n,+} \rangle \langle \epsilon_{n+1,-}\vert )  \nonumber \\
&& + \tilde{\gamma}_{1}s^{2}_{0}\hat{\mathcal{D}} (\vert\epsilon_{0,+} \rangle \langle \epsilon_{0}\vert )
+\tilde{\gamma}_{2}c^{2}_{0}\hat{\mathcal{D}} (\vert\epsilon_{0,-} \rangle \langle \epsilon_{0}\vert )\nonumber\\
&&+ \sum_{n=0}^{\infty}\tilde{\gamma}_{3}a_{n}^2 \hat{\mathcal{D}} (\vert \epsilon_{n+1,+} \rangle \langle \epsilon_{n,+}\vert )+
\sum_{n=0}^{\infty}\tilde{\gamma}_{4}b_{n}^2 \hat{\mathcal{D}} (\vert \epsilon_{n+1,-} \rangle \langle \epsilon_{n,-}\vert ) \nonumber \\
&& + \sum_{n=0}^{\infty}\tilde{\gamma}_{5}d_{n}^2 \hat{\mathcal{D}} (\vert \epsilon_{n+1,+} \rangle \langle \epsilon_{n,-}\vert ) 
+ \sum_{n=0}^{\infty}\tilde{\gamma}_{6}d_{n}^2 \hat{\mathcal{D}} (\vert \epsilon_{n+1,-} \rangle \langle \epsilon_{n,+}\vert ), 
\end{eqnarray}
where we have used the standard notation for dissipators $\hat{\mathcal{D}}(\hat{O}) = \hat{O} \rho \hat{O}^{\dagger}-
\{ \hat{O}^{\dagger}  \hat{O},\rho\} / 2$. The auxiliary coefficients are defined trough the relations,
\begin{eqnarray}
a_n=c_{n}c_{n+1}\sqrt{n+1} ~+~ s_{n}s_{n+1}\sqrt{n+2}, \nonumber \\ b_n=s_{n}s_{n+1}\sqrt{n+1} ~+~ c_{n}c_{n+1}\sqrt{n+2}, \nonumber \\
d_n=s_{n}c_{n+1}\sqrt{n+2} ~-~ c_{n}s_{n+1}\sqrt{n+1},
\end{eqnarray} and the explicit frequency-dependent decay rates are given by,
\begin{eqnarray}
\begin{array}{lcllcl}
\gamma_1 &=& \left[1+\bar{n}(\frac{\omega_0+\omega+\Omega_0}{2})\right] \gamma,
&\tilde{\gamma}_1 &=& \bar{n}(\frac{\omega_0+\omega+\Omega_0}{2}) \gamma,  \\ 
\gamma_2&=& \left[1+\bar{n}(\frac{\omega_0+\omega-\Omega_0}{2})\right] \gamma,
& \tilde{\gamma}_2 &=& \bar{n}(\frac{\omega_0+\omega-\Omega_0}{2}) \gamma, \\ 
\gamma_3 &=& \left[1+\bar{n}(\omega+\frac{\Omega_{n+1}-\Omega_{n}}{2})\right] \gamma, 
& \tilde{\gamma}_3 &=& \bar{n}(\omega+\frac{\Omega_{n+1}-\Omega_{n}}{2}) \gamma,\\ 
\gamma_4 &=& \left[1+\bar{n}(\omega-\frac{\Omega_{n+1}-\Omega_{n}}{2})\right] \gamma,
&\tilde{\gamma}_4 &=& \bar{n}(\omega-\frac{\Omega_{n+1}-\Omega_{n}}{2}) \gamma,  \\
\gamma_5 &=& \left[1+\bar{n}(\omega+\frac{\Omega_{n+1}+\Omega_{n}}{2})\right] \gamma ,
&\tilde{\gamma}_5 &=& \bar{n}(\omega+\frac{\Omega_{n+1}+\Omega_{n}}{2}) \gamma, \\
\gamma_6 &=& \left[1+\bar{n}(\omega-\frac{\Omega_{n+1}+\Omega_{n}}{2})\right] \gamma,
&\tilde{\gamma}_6 &=&  \bar{n}(\omega-\frac{\Omega_{n+1}+\Omega_{n}}{2}) \gamma,
\end{array}
\end{eqnarray} 
where we have used the Kubo-Martin-Schwinger condition to relate the downward and upward transition rates,
$\tilde\gamma(\nu)=\exp\left(-\nu/k_{B} T\right)\gamma(\nu)$.
Figure \ref{fig:Fig2} shows these decay channels in the dressed state ladder of the JC model.
Note that the resonant model, $\Delta=0$, reduces to the one obtained in Ref.~\cite{Scala2007p013811} in the zero-$T$ limit, $T\rightarrow 0$, $\bar{n}(\nu)\rightarrow 0$, 
and $\gamma_{i}\rightarrow\gamma$, $\tilde\gamma_{i}\rightarrow 0$.

In the following, we compare the microscopic approach with the standard phenomenological approach for field dissipation with a flat environment, 
which is commonly described in the literature by the following master equation \cite{Haroche1992, CohenTannoudji1998},
\begin{eqnarray}\label{eq:phME}
\dot\rho_{\rm{ph}}(t)&=&-{ i} [\hat H_{JC},\rho_{\rm{ph}}(t)]+ \gamma \left[\bar{n}(\omega)+1 \right] \left[ \hat{a} \rho_{\rm{ph}}(t) \hat{a}^{\dagger} - \frac{1}{2} \{\hat{a}^{\dagger} \hat{a}, \rho_{\rm{ph}}(t)\} \right] \nonumber \\
&&+ \gamma \bar{n}(\omega) \left[ \hat{a}^{\dagger}\rho_{\rm{ph}}(t) \hat{a} - \frac{1}{2} \{\hat{a} \hat{a}^{\dagger},\rho_{\rm{ph}}(t)\} \right],
\end{eqnarray}
and is valid for a broader range of parameters provided that coupling is small compared to the free field and qubit frequencies, $\omega,\omega_0\gg\gamma$.

\section{Single-excitation manifold at zero temperature}
\begin{figure}
	\centering
	\includegraphics{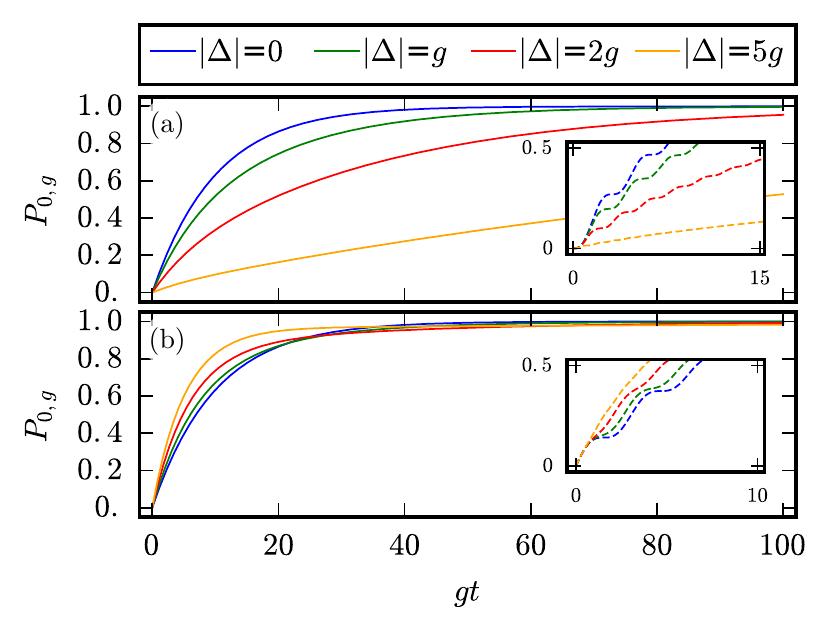}	
	\caption{Probability of finding the system in the ground state for initial states (a) $\vert \psi(0) \rangle = \vert 0,e \rangle$ and (b) $\vert \psi(0) \rangle = \vert 1,g \rangle$ under the dynamics provided by the microscopic description of dissipation at zero-$T$ and different detunings. Insets: phenomenological description. Simulation parameters: $\{\gamma,\omega_0\}=\{~0.2,~100\}g$.} \label{fig:Fig3}
\end{figure}

One of the basic signatures in  the dynamics of the JC model are the so called Rabi oscillations, showing the periodic exchange of excitations between the qubit and the field mode. In absence of losses, this is an indefinitely reversible process of coherent evolution. 
In real cavity-QED experiments \cite{Brune1996p1800}, the cavity is in fact open and subject to decoherence, making Rabi oscillations decay and eventually disappear with the inevitable scape of the photon to the environment.  
Single-excitation dynamics at zero temperature under the microscopic approach \cite{Scala2007p013811} are described by the following simplified form of Eq. (\ref{eq:DSME}),
\begin{eqnarray}
\dot{\rho}(t)&=& - i [\hat H_{JC},\rho(t)] + \gamma\left[ s^{2}_{0} \hat{\mathcal{D}} (\vert \epsilon_{0} \rangle \langle \epsilon_{0,+}\vert )
+ c^{2}_{0}\hat{\mathcal{D}} (\vert\epsilon_{0} \rangle \langle \epsilon_{0,-}\vert )\right],
\end{eqnarray}
which immediately shows  jump operators describing transitions from states $\vert\epsilon_{0,\pm}\rangle$ to the ground state $\vert\epsilon_0\rangle$. 
In fact, in this microscopic description with dressed states, decay of the two bare states $\vert e,0\rangle$ and $\vert{g,1}\rangle$ is allowed, in contrast to 
the  phenomenological description where  the bare state $\vert{g,1}\rangle$ provides the only decay channel to the ground sate. 
Furthermore, it is possible to construct an analytic solution for the case of pure initial states in the single-excitations manifold, $\vert \psi(0)\rangle= \alpha \vert 0,e\rangle + \beta \vert 1,g\rangle$ with $\beta =  \sqrt{1- \vert \alpha^2 \vert}$, using the damping basis technique  \cite{Briegel1993p3311}
\begin{eqnarray}
\rho(t)&=& \left\{ 1 - \left[ \vert \alpha \vert^{2} +c_{0}^2- 2 c_{0} s_{0} \Re\left( \alpha \beta^{\ast} \right) \right] e^{-\gamma s_{0}^2 t} + \right. \nonumber \\
&&\left. - \left[ \vert \alpha \vert^{2} + s_{0}^2 + 2 c_{0} s_{0} \Re\left(\alpha\beta^{\ast}\right)\right] e^{-\gamma c_{0}^2 t} \right\}
\vert\epsilon_{0} \rangle \langle  \epsilon_{0}\vert + \nonumber\\
&& + \left[ \vert\alpha\vert^{2} + c_{0}^2 - 2 c_{0} s_{0} \Re\left(\alpha \beta^{\ast}\right)\right]e^{-\gamma s_{0}^2 t}
\vert\epsilon_{0,-} \rangle \langle  \epsilon_{0,-}\vert + \nonumber \\
&& + \left[ \vert \alpha \vert^{2} + s_{0}^2 + 2 c_{0} s_{0} \Re\left(\alpha\beta^{\ast}\right) \right] e^{-\gamma c_{0}^2 t}
\vert\epsilon_{0,+} \rangle \langle  \epsilon_{0,+}\vert + \nonumber \\
&&+ e^{-\gamma t/2} \left\{ \left[ \left( 1 - 2 \vert \alpha \vert^{2} \right) c_{0} s_{0} + \alpha^{*} \beta c_{0}^{2}
-\alpha \beta^{*} s_{0}^{2} \right] e^{i\Omega_{0}t} \vert\epsilon_{0,-} \rangle \langle  \epsilon_{0,+}\vert + \right. \nonumber \\
&& \left. + \left[ \left( 1 - 2 \vert \alpha \vert^{2} \right) c_{0} s_{0} + \alpha \beta^{*} c_{0}^{2}
-\alpha^{*} \beta s_{0}^{2} \right] e^{-i\Omega_{0}t} \vert\epsilon_{0,+} \rangle \langle  \epsilon_{0,-}\vert \right\}. 
\end{eqnarray}
Off-resonant interaction makes one of the two decay channels dominant, and gives the possibility to control the decay to the ground state; for example, as  we increase the detuning
of the qubit-field interaction, the coherent exchange of the excitation is maintained for longer times or, equivalently, the life time of the photon inside the cavity increases as we can see in Fig. \ref{fig:Fig3}. 
This interesting asymmetry could be useful for increasing  the number of operations trough simple quantum gates using cavity-QED implementations. 

Meanwhile, the phenomenological description in the single-excitation manifold, 
\begin{eqnarray}
\dot\rho_{\rm{ph}}(t)&=&- i [\hat H_{JC},\rho_{\rm{ph}}(t)]+ \gamma \hat{\mathcal{D}} (\vert 0,g \rangle \langle 1,g \vert ),
\end{eqnarray}
shows the direct decay of the state $\vert 1,g \rangle$ to the ground state. For this master equation, it is also possible to find an exact solution for the same initial state as before,
\begin{eqnarray}
\rho_{\rm{ph}}(t)&=& \left[ 1 - \vert a(t) \vert^{2} - \vert b(t) \vert^{2} \right] \vert \epsilon_{0} \rangle \langle \epsilon_{0} \vert +  \vert\Psi(t)\rangle\langle\Psi(t)\vert  ,
\end{eqnarray}
where the time evolution of the single-excitation state $\vert \Psi(t)\rangle$ is given by
\begin{eqnarray}
\vert\Psi(t)\rangle=\left[ c_{0} a(t) + s_{0}  b(t) \right] \vert \epsilon_{0,+} \rangle + \left[c_{0} b(t) - s_{0} a(t) \right] \vert \epsilon_{0,-} \rangle,
\end{eqnarray} 
with the time-dependent functions,
\begin{eqnarray}
a(t)&=&\left[ \alpha \cosh \frac{ \Omega t}{2} +\frac{\alpha\left(\gamma-2i\Delta\right)-4i\beta g}{2 \Omega}\sinh \frac{ \Omega t}{2} \right]
e^{-\frac{1}{4}\left(\gamma+6i\Delta\right)t }, \nonumber \\
b(t) &=& \left[\beta\cosh\frac{ \Omega t}{2} -\frac{\beta\left(\gamma-2i\Delta\right)+4i\alpha g}{2\tilde\Omega}\sinh\frac{ \Omega t}{2}\right]e^{-\frac{1}{4}\left(\gamma+6i\Delta\right)t },
\end{eqnarray}
where the auxiliar frequency $\Omega=\sqrt{\gamma^2-16g^2-4\Delta\left(\Delta+i\gamma\right)}/2$ is complex. 
The only difference between the two treatments is the presence of a high-frequency modulation, at short propagation times, in the decay to the ground state dynamics under the phenomenological description, insets Fig. \ref{fig:Fig3}.

\begin{figure}
	\centering
	\includegraphics{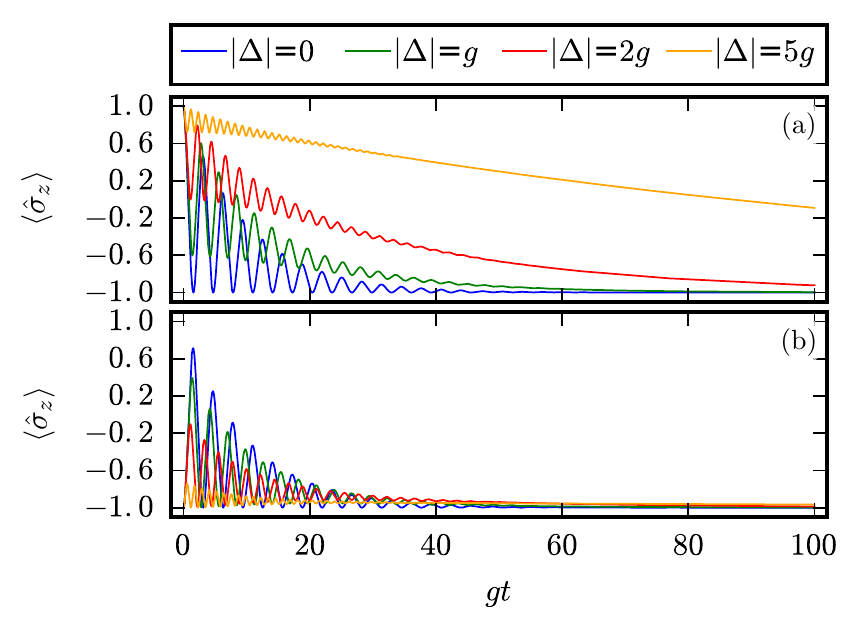}	
	\caption{Evolution of the population  inversion for initial states (a) $\vert \psi(0) \rangle = \vert 0,e \rangle$ and (b) $\vert \psi(0) \rangle = \vert 1,g \rangle$ under the dynamics provided by the microscopic description of dissipation at zero-$T$ and different detunings. Simulation parameters: $\{\gamma,\omega_0\}=\{~0.2,~100\}g$.}\label{fig:Fig4}
\end{figure}

\begin{figure}
	\centering
	\includegraphics{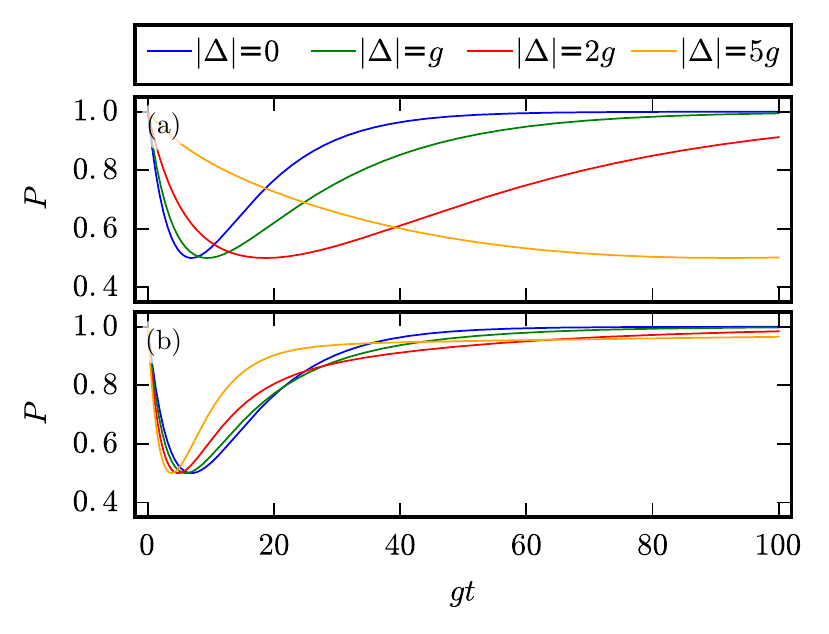}	
	\caption{Evolution of the qubit-field purity for initial states (a) $\vert \psi(0) \rangle = \vert 0,e \rangle$ and (b) $\vert \psi(0) \rangle = \vert 1,g \rangle$ under the dynamics provided by the microscopic description of dissipation at zero-$T$ and different detunings. Simulation parameters: $\{\gamma,\omega_0\}=\{~0.2,~100\}g$.}\label{fig:Fig5}
\end{figure}

\begin{figure}
	\centering
	\includegraphics{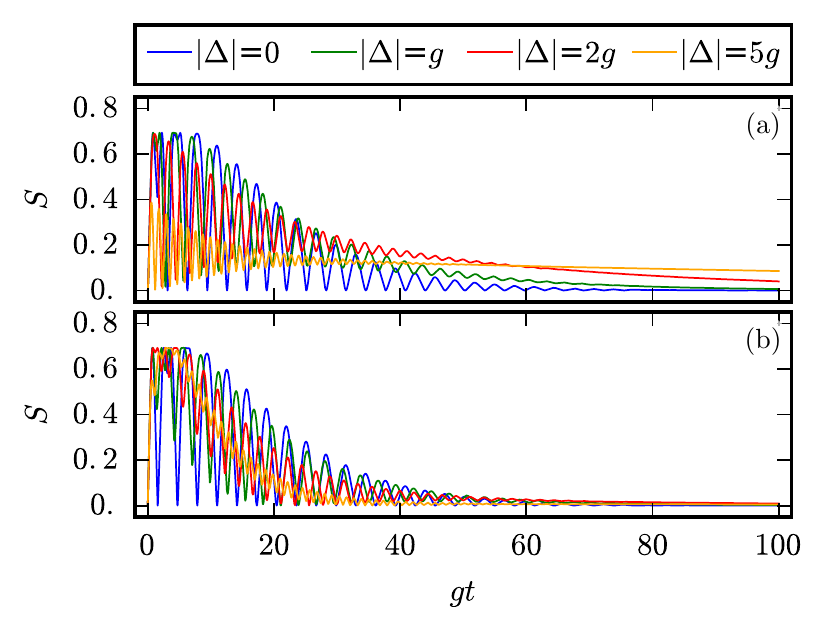}	
	\caption{Evolution of von Neumann entropy of the field for initial states (a) $\vert \psi(0) \rangle = \vert 0,e \rangle$ and (b) $\vert \psi(0) \rangle = \vert 1,g \rangle$ under the dynamics provided by the microscopic description of dissipation at zero-$T$ and different detunings. Simulation parameters: $\{\gamma,\omega_0\}=\{~0.2,~100\}g$.}\label{fig:Fig6}
\end{figure}

Figure \ref{fig:Fig3} shows the probability to find the system in the ground state for a near-resonance system, $\omega_{0} \sim \omega \gg g$, for different detuning between the qubit and field frequencies for initial states in the single-excited state manifold. 
An initial qubit in the excited state, $\vert \psi(0) \rangle = \vert 0,e \rangle$, produces slower effective decay to the ground state with larger absolute values of the detuning, Fig. \ref{fig:Fig3}(a), while an initial qubit in the ground state, $\vert \psi(0) \rangle = \vert 0,g \rangle$, produces larger effective decay rates to the ground state with larger absolute values of the detuning, Fig. \ref{fig:Fig3}(b).
The same process is observed in the phenomenological approach with the addition of a higher frequency oscillation, insets in Fig. \ref{fig:Fig3}.
The damped Rabi oscillations in the atomic inversion dynamics, $\langle \hat{\sigma}_{z} \rangle = \mathrm{Tr}\{\sigma_{z}\rho_{q}(t)\}$, can be seen in Figure \ref{fig:Fig4}. 
Regardless of the effective decay rate to the ground state, the resonant system will go faster to a pure state, as shown by the qubit-field  purity, $ P = \mathrm{Tr}~ \rho^2(t)$ in Fig. \ref{fig:Fig5}, and von Neumann entropy for the field, $S = -\mathrm{Tr}~ \rho_{f}(t) \ln \rho_{f}(t)$ in Fig. \ref{fig:Fig6}.
Purity minima appears at longer scaled times for larger absolute values of the detuning for the initial state  $\vert \psi(0) \rangle = \vert 0,e \rangle$, Fig. \ref{fig:Fig5}(a), and the opposite happens for the initial state $\vert \psi(0) \rangle = \vert 1,g \rangle$, Fig. \ref{fig:Fig5}(b). 
On-resonance, $\Delta = 0$, and for an initial state $\vert \psi(0) \rangle = \vert 0,e \rangle $, it is possible to find a simple expression for the purity, $ P(t)=1-2\left[1-e^{-\gamma t/2}\right]e^{-\gamma t /2}$, that reaches its minimum at the scaled time $gt = 2 \ln 2/\gamma$.

In the case of the closed JC model, the von Neumann entropy of the field provides a measure of the entanglement between the qubit and field \cite{Gerry2005}. However, this is not true 
if the system is open. Here, it is not possible to express the qubit-field state vector at any time using the appropriate Schmidt decomposition,  hence
the respective qubit and field von Neumann entropies are not expected to be equivalents giving rise to different behaviors. Actually, the information flow from 
the qubit-field system to the environment must be reflected in the entropies of the subsystems. 
In Figure \ref{fig:Fig6}, we show  the von Neumann entropy of the field as a function of time for the two initial bare states in the single excitation manifold, showing the usual dynamics of entanglement and disentanglement of the qubit-field system but with the effects of damping, or decoherence, due to the interaction with the environment.
It provides a highly oscillatory picture, that depends on the detuning, for how the time evolution of the field reduced density matrix departures from, and asymptotically comes back, to a quantum pure state. 

\begin{figure}
	\centering
	\includegraphics{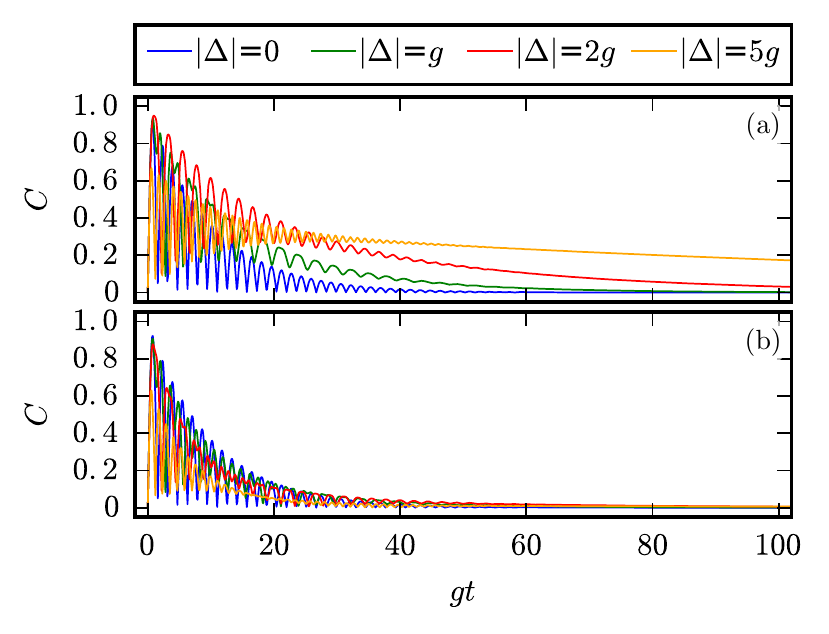}	
	\caption{Evolution of the qubit and single-excitation field concurrence for initial states (a) $\vert \psi(0) \rangle = \vert 0,e \rangle$ and (b) $\vert \psi(0) \rangle = \vert 1,g \rangle$ under the dynamics provided by the microscopic description of dissipation at zero-$T$ and different detunings. Simulation parameters: $\{\gamma,\omega_0\}=\{~0.2,~100\}g$.}\label{fig:Fig7}
\end{figure}

In the single-excitation limit, we can think of the field  as an effective qubit, and calculate the two-qubit concurrence for the field-matter state.
The concurrence, an entanglement measure based on the concept of entanglement of formation \cite{Wootters1998p2245,Wootters2001p27}, is defined as $C = 2\mathrm{max}\{\lambda_i\}-\sum_{i}\lambda_i$, where the set $\{\lambda_{i}\}$ are the square roots of the eigenvalues of the operator  $R=\rho\left(\sigma_y\otimes\sigma_y\right)\rho^{*}\left(\sigma_y\otimes\sigma_y\right)$ in decreasing order.
The concurrence is zero for separable states, and takes its maximum value of one for maximally entangled states.
Its behaviour under the microscopic approach shows the evolution from the separable initial state to an almost maximally entangled state in the time for half a Rabi oscillation in the single-excitation manifold, Fig. \ref{fig:Fig7}. Obviously, this will be affected by the decoherence induced by the environment. The effective decay rates induced show that an initial state $\vert \psi(0) \rangle = \vert 0,e \rangle$ produces a higher entangled state for small evolution times and higher detuning, Fig. \ref{fig:Fig7}(a).
An initial pure separable state of the form $\vert \psi(0) \rangle = \vert 1,g \rangle$  produces and maintains higher concurrence values for lower detuning, Fig. \ref{fig:Fig7}(b). This is
in agreement with the information provided by von Neumann entropy and our previous discussion on the effect of the detuning on the effective decay rates.

\section{Beyond the single-excitation manifold at finite temperature}

As we go beyond the single-excitation manifold, starting with an initial state with more than one total excitation, the oscillations in the ground state probability, $P_{0,g}$, provided by dynamics in the phenomenological description, insets in Fig. \ref{fig:Fig3}, have larger frequencies and, eventually, make the  the phenomenological description indistinguishable to the naked eye from that of the microscopic one using the variables presented above.
Here, we will show that it is possible to use phase space dynamics to notice the differences between the two approaches.
Sadly, it becomes cumbersome  and impractical to address analytically the dynamics beyond the single-excitation manifold at zero-$T$, and we must resort to numeric simulations in order to create intuition for these systems. 
In the following, we numerically solve the microscopic master equation, Eq. (\ref{eq:DSME}), by two methods, brute force iterative Runge--Kutta methods and direct diagonalization of the Liouvillian \cite{NavarreteBenlloch2015}, in both cases the dimension of the master equation is truncated once a desired convergence is reached.

\subsection{Fock states}

\begin{figure}
	\centering
	\includegraphics{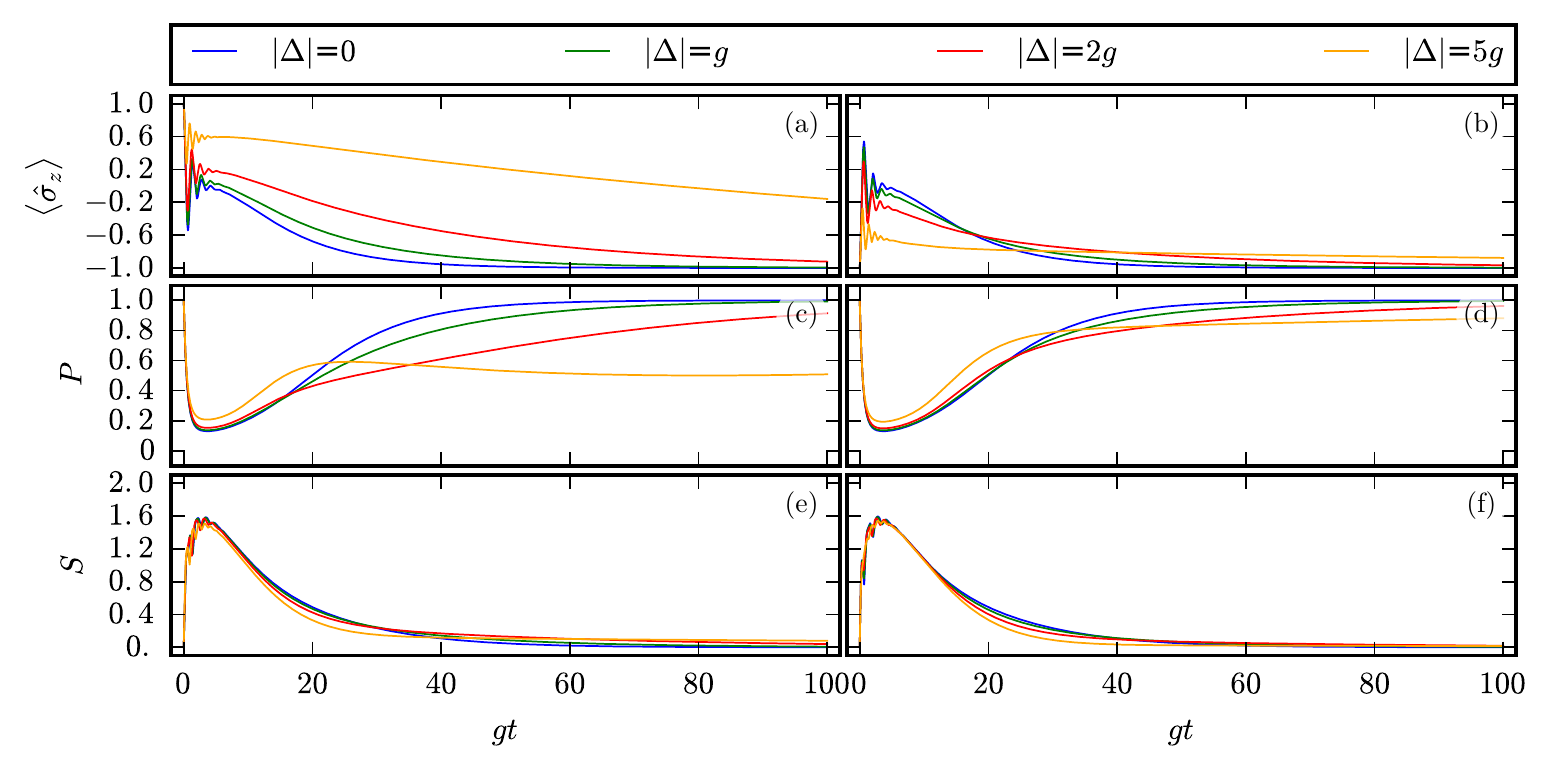}	
	\caption{Time evolution of the atomic inversion (first row), qubit-field purity (second row), and von Neumann entropy of the field (third row), for initial states $\vert \psi(0) \rangle = \vert 4,e \rangle$ (left column) and $\vert \psi(0) \rangle = \vert 5,g \rangle$ (right column), under dynamics provided by the microscopic approach to dissipation at zero-$T$ for different detuning between the qubit and field frequencies. Simulation parameters: $\{\gamma,\omega_0\}=\{~0.2,~100\}g$} \label{fig:Fig8}
\end{figure}

At zero-$T$, an initial state in the $\langle N\rangle$-excitation manifold, $\vert \psi(0) \rangle = \vert n,e \rangle$ or $\vert \psi(0) \rangle = \vert n+1,g \rangle$, should present similar dissipation dynamics to those described above: the effective decay rate for initial excited and ground state dynamics will differ and be related to the detuning between the qubit and field frequencies.
We can see this in the time evolution of the atomic inversion for an initial state $\vert \psi(0) \rangle = \vert 4,e \rangle$, Fig. \ref{fig:Fig8}(a), and $\vert \psi(0) \rangle = \vert 5,g \rangle$, Fig. \ref{fig:Fig8}(b), but becomes more evident in the qubit-field purity, Fig. \ref{fig:Fig8}(c) and Fig. \ref{fig:Fig8}(d), and von Neumann entropy of the field, Fig. \ref{fig:Fig8}(e) and Fig. \ref{fig:Fig8}(f). 
The dynamics provided by the phenomenological approach still have a higher modulating frequency, but it becomes so high that the differences are indistinguishable without further analysis.

\begin{figure}
	\centering
	\includegraphics{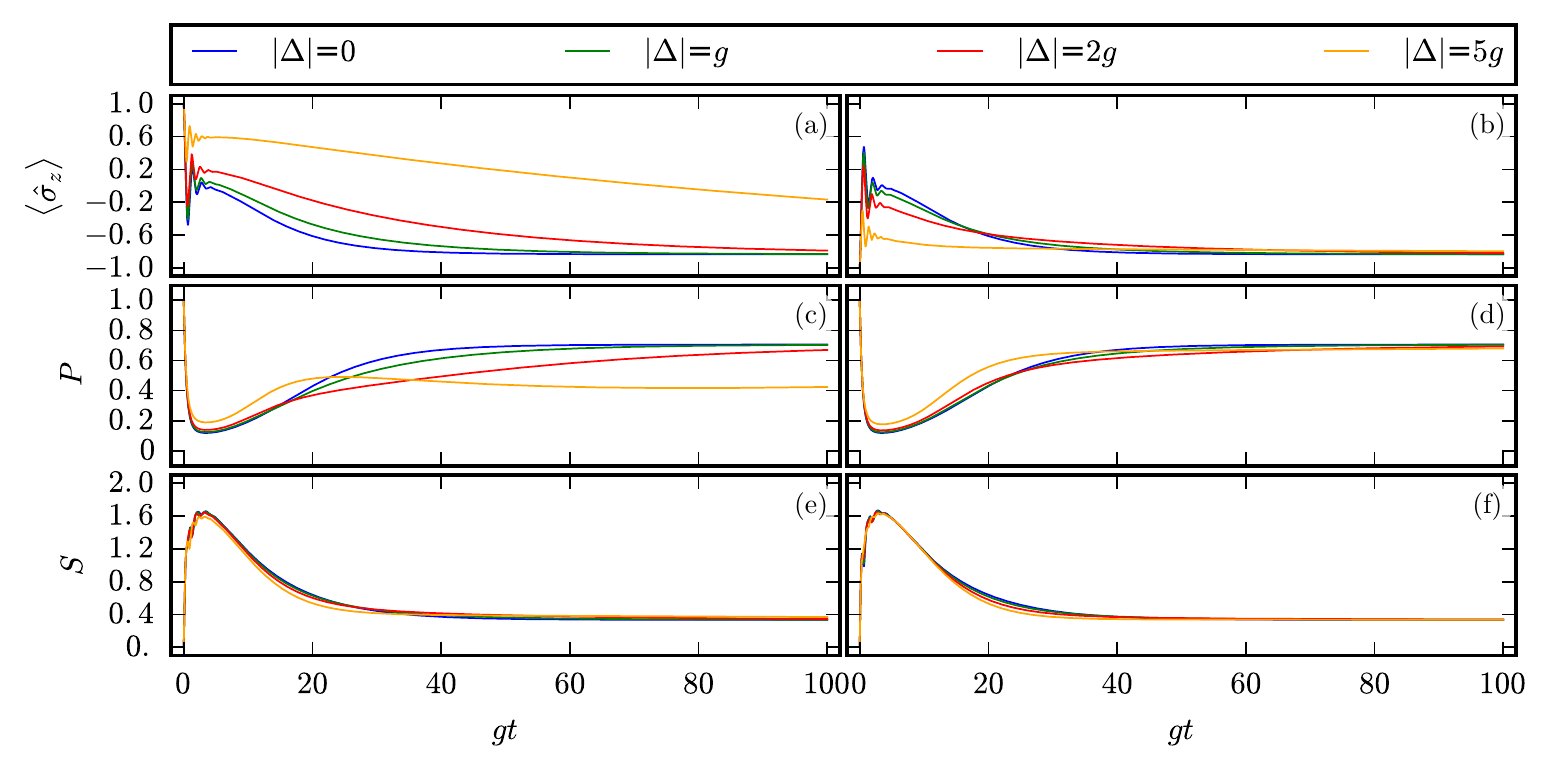}	
	\caption{Time evolution of the atomic inversion (first row), qubit purity (second row), and von Neumann entropy of the field (third row), for initial states $\vert \psi(0) \rangle = \vert 4,e \rangle$ (left column) and $\vert \psi(0) \rangle = \vert 5,g \rangle$ (right column), under dynamics provided by the microscopic approach to dissipation at low-$T$, for different detuning between the qubit and field frequencies. Simulation parameters: $\{\gamma,\omega_0,\bar{n}\}=\{0.2g,~100g,~0.1\}$.} \label{fig:Fig9}
\end{figure}

At finite-$T$, the dynamics are equivalent to those at zero-$T$ with a slight increase of the effective decay rate due to temperature effects and, obviously, the final state of the radiation-matter system, in the asymptotic limit, will reach the thermal equilibrium steady state of the open system.
Figure \ref{fig:Fig9}(a) and \ref{fig:Fig9}(b) shows the time evolution of the atomic inversion, Fig. \ref{fig:Fig9}(c) and \ref{fig:Fig9}(d) that of the qubit-field purity, and Fig. \ref{fig:Fig9}(e) and \ref{fig:Fig9}(f) the time evolution of the von Neumann entropy of the field of initial states $\vert \psi(0) \rangle = \vert 4,e \rangle$ and $\vert \psi(0) \rangle = \vert 5,g \rangle$, in that order for each case, under JC dynamics interacting with a low-$T$ thermal environment with average thermal photons $\bar{n}=0.1$; a value related to cavity-QED experiments \cite{Wilczewski2009p033836,Wilczewski2009p013802}.

\subsection{Coherent states}

\begin{figure}
	\centering
	\includegraphics{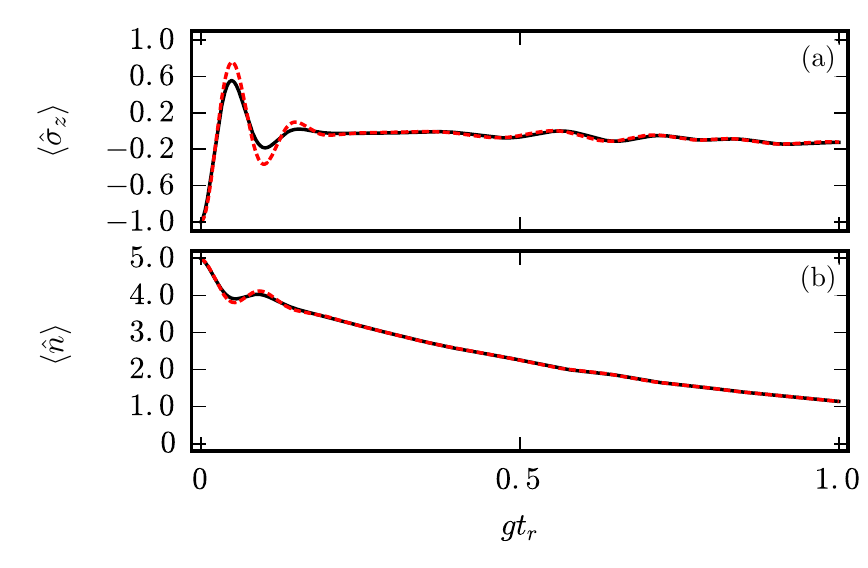}
	\caption{Time evolution of the (a) atomic inversion and (b) mean photon number for an initial state composed of a coherent field with $\alpha = \sqrt{5}$ and the atom in the ground state on-resonance, $\Delta=0$, under dynamics ruled by microscopic (black solid lines) and phenomenological (dotted red lines) approaches to dissipation. Simulation parameters: $\{\gamma,\omega_0,\bar{n}\}=\{0.1g,~100g,~0\}$. }\label{fig:Fig10}
\end{figure}

In order to study more complex dynamics, let us consider initial states involving coherent states of the field, $\vert\alpha\rangle= \exp(-|\alpha|^{2}/2) \sum_{n=0}^{\infty} \alpha^{n}/\sqrt{n!} \vert n\rangle$.
These are the most classical quantum states in which a field mode can be prepared, thus the name of semi-classical states of the field.
For the sake of simplicity, we start from a pure and separable initial state, $\vert \psi(0) = \vert \alpha, g \rangle$, that shows collapse and revival of the atomic inversion at the approximate scaled revival time $g t_{r} \sim 2\pi \sqrt{\vert \alpha \vert^2}$ for the closed system.
Figure \ref{fig:Fig10} shows the atomic inversion and mean photon number evolution under the microscopic and phenomenological approaches to dissipation for a single revival time.
Cavity losses slightly affect the initial collapse of the atomic inversion, but heavily suppress the revival, Fig \ref{fig:Fig10}(a), in agreement with previous results employing the phenomenological approach.
It is possible to observe differences between the two approaches at short times, but the dynamics seem to become identical as the system evolves.
Note that care must be exerted to use simulation parameters that satisfy the restrictions mentioned above for each model.

\begin{figure}
	\centering
	\includegraphics{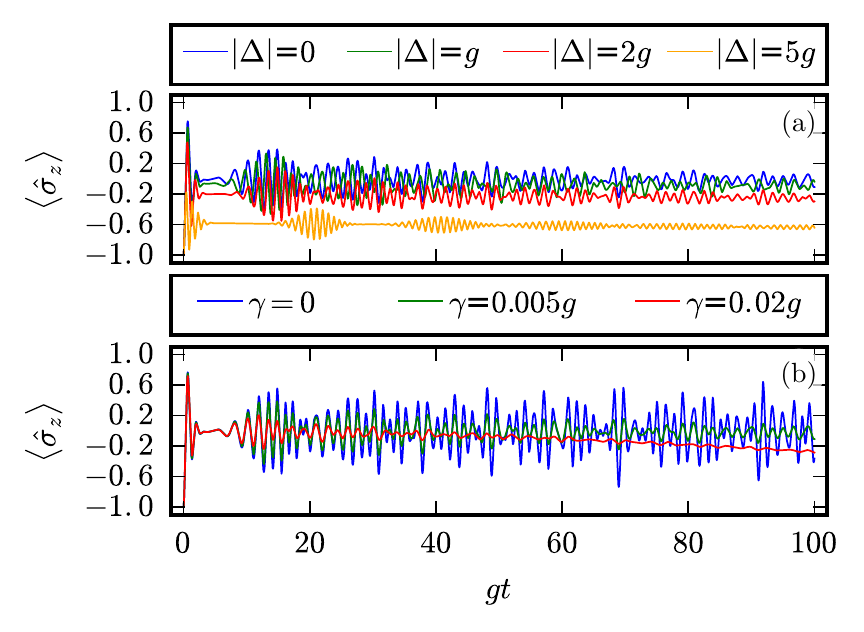}
	\caption{Time evolution of the atomic inversion for an initial state composed of a coherent field with $\alpha = \sqrt{5}$ and the atom in the ground state, $\vert \psi(0) \rangle = \vert \alpha, g \rangle$. (a) Varible detuning with fixed decay rate $\gamma$, simulation parameters: $\{\gamma,\omega_0\}=\{0.005,~100\}g$, (b) variable decay rate on-resonance, simulation parameters: $\omega_0= \omega = 100g$.  }\label{fig:Fig11}
\end{figure}

Obviously, the effects of detuning at each and every manifold with constant total excitation number described above will survive. 
For example, an initial state composed by a coherent field and the two-level system in the ground state, $\vert \psi(0) \rangle = \vert \alpha, g \rangle$, will have a lower effective decay rate for larger detuning, Fig. \ref{fig:Fig11}(a), as expected.
Figure \ref{fig:Fig11}(b) shows the atomic inversion evolution, on-resonance for different decay rates, where we can observe that the collapse dynamics, for times shorter than half the revival time, are barely modified while the revival dynamics is strongly suppressed for increasing decay rate.

\begin{figure}
	\centering
	\includegraphics{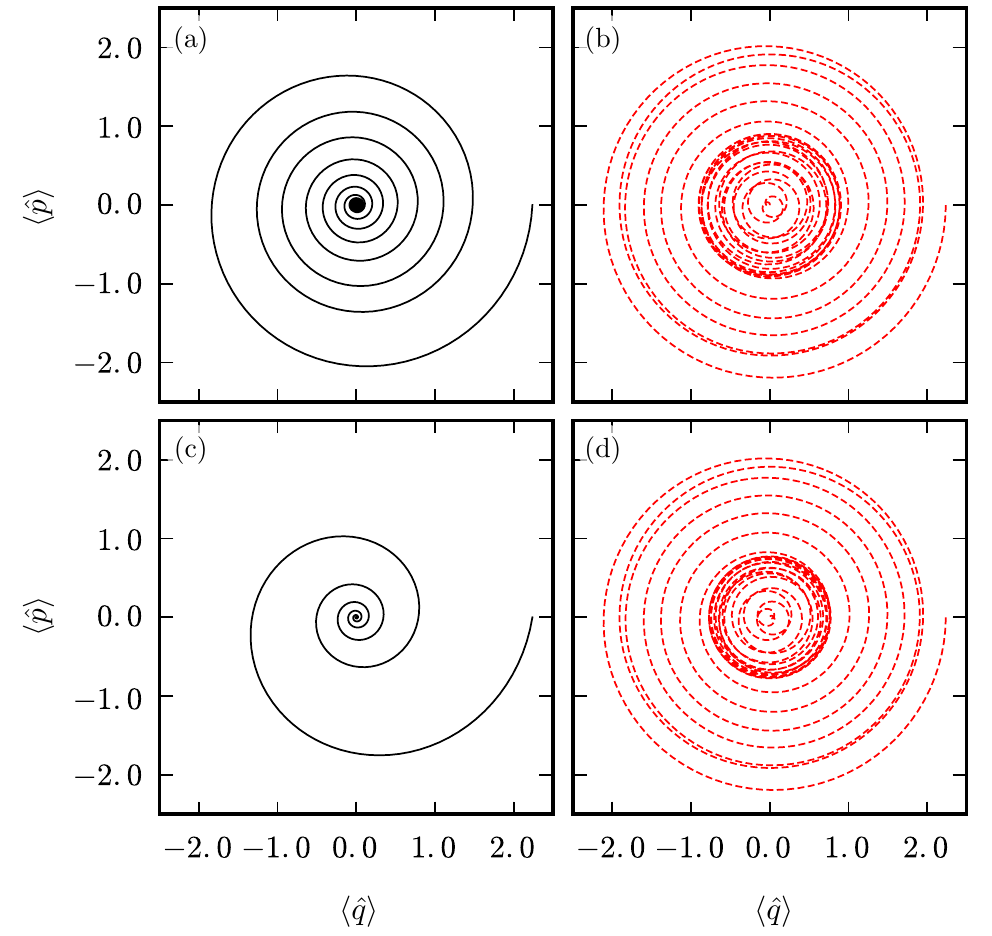}
	\caption{Time evolution of the mean value of the field quadratures for an initial state $\vert \psi(0) \rangle = \vert \alpha, g \rangle$ with $\alpha = \sqrt{5}$ under the microscopic (left column) and phenomenological approaches (right column) to dissipation at zero-$T$ (first row), simulation parameters: $\{\gamma,\omega_0\}=\{0.1,~100\}g$, and finite-$T$ (second row), simulation parameters: $\{\gamma,\omega_0,\bar{n}\}=\{0.1g,~100g,~1.0\}$. All cases consider a simulation scaled time interval $[0,2 g t_{r}]$.}\label{fig:Fig12}
\end{figure}

A substantial deviation between the two approaches is easier to detect using the time evolution of the field quadratures, $\hat{q} = \left( \hat{a} + \hat{a}^{\dagger} \right)/2$ and $\hat{p} = \left( \hat{a} -\hat{a}^{\dagger}\right)/2i $, whose mean values for a coherent state are equivalent to the real and imaginary part of the analogue classical complex field amplitude. 
Interestingly enough, the microscopic approach to dissipation provides us with an intuitively expected, spiral decay evolution of the field quadratures, Fig. \ref{fig:Fig12}(a), similar to the one obtained by the phenomenological approach for just a dissipative cavity. 
The time evolution for the field quadratures under the phenomenological approach shows the differences and high frequency modulation in the form of deviations from the spiral decay of the free dissipative field, Fig. \ref{fig:Fig12}(b).
Furthermore, the effect of finite-$T$, an increased decay rate, is more evident in the microscopic approach, Fig. \ref{fig:Fig12}(c), than in the phenomenological approach, Fig.\ref{fig:Fig12}(d), both in short- and moderate-time scales.

\begin{figure}
	\centering
	\includegraphics{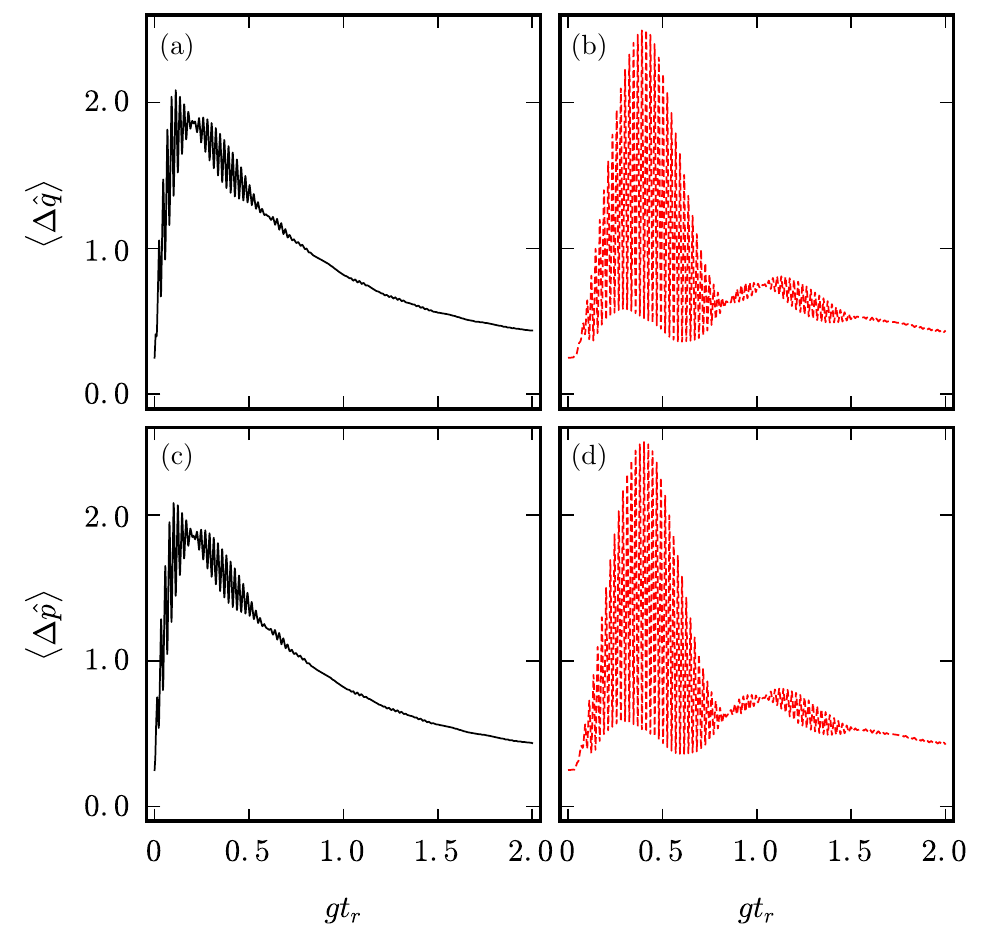}
	\caption{Time evolution of the field quadratures variances for an initial state $\vert \psi(0) \rangle = \vert \alpha, g \rangle$ with $\alpha = \sqrt{5}$ under the microscopic (left column) and phenomenological approaches (right column) to dissipation at zero-$T$, simulation parameters: $\{\gamma,\omega_0\}=\{0.1,~100\}g$.}\label{fig:Fig13}
\end{figure}

The variances of the field quadratures, $\langle \Delta \hat{x} \rangle = \langle \hat{x}^2 \rangle - \langle \hat{x} \rangle^2$, for the microscopic, Fig. \ref{fig:Fig13}(a) and Fig. \ref{fig:Fig13}(c), and the phenomenological approaches, Fig. \ref{fig:Fig13}(b) and Fig. \ref{fig:Fig13}(d), show an even greater difference on the open system dynamics provided by the two approaches.
Under the microscopic approach to dissipation, the initial coherent state of the field stops minimizing the uncertainty relation for the field quadratures in a shorter time than under phenomenological open dynamics.
Furthermore, open microscopic dynamics predict lower fluctuations in the variances of the field quadratures, leading to a smoother transition to the steady state, the coherent vacuum state at zero-$T$, than the one predicted by phenomenological open dynamics.

\begin{figure}
	\centering
	\includegraphics{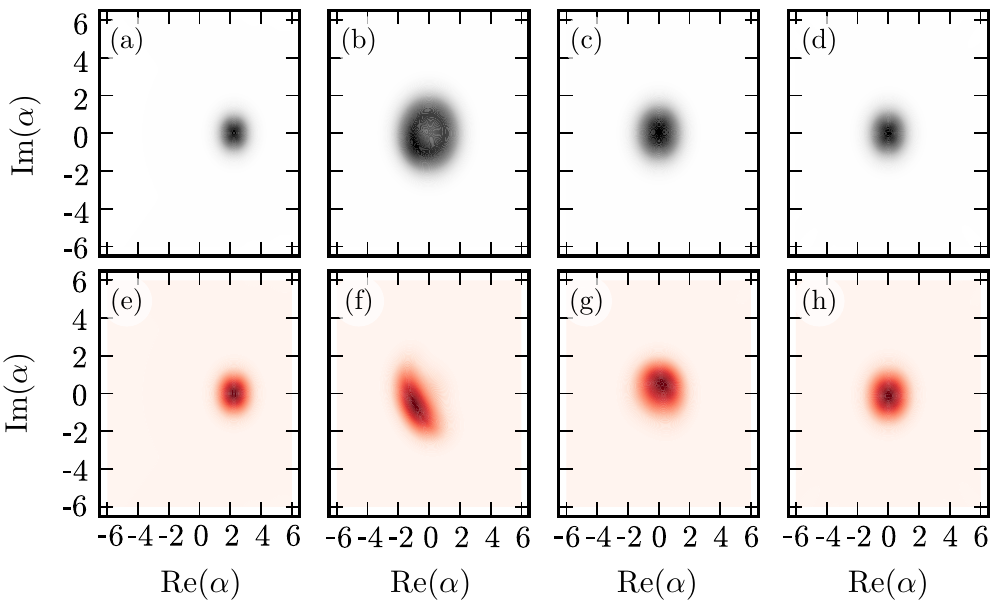}
	\caption{Snapshots of Husimi Q-function for different evolution times of an initial state $\vert \psi(0) \rangle = \vert \alpha, g \rangle$ with $\alpha = \sqrt{5}$ under the microscopic (top row) and phenomenological approaches (bottom row) to dissipation at zero-$T$ at scaled times (a),(e) $gt= 0$, (b),(f) $gt= 2g t_{r} / 3$, (c),(g) $gt= 4g t_{r}/3 $, and (d),(h) $gt= 2g t_{r}$ with simulation parameters: $\{\gamma,\omega_0\}=\{0.1,~100\}g$.}\label{fig:Fig14}
\end{figure}

These differences in the mean values of the quadratures and their variances can also be observed in phase space thorough quasi-probability distributions, like Husimi Q-function, $Q(\alpha)=  \langle \alpha \vert \hat{\rho} \vert \alpha \rangle / \pi$, shown in Fig. \ref{fig:Fig14}.
The dynamics of the Q-function under a microscopic description of dissipation starts from a well defined Gaussian phase space distribution corresponding to a coherent state that, smoothly and quickly, becomes a donut-shaped distribution whose radius starts diminishing until it takes the Gaussian distribution form of coherent vacuum. Fig. \ref{fig:Fig14}(a)-Fig. \ref{fig:Fig14}(d).
Meanwhile, the evolution of the Q-function under the phenomenological description follows a more complicated dynamics that might look like a decaying spiral to the coherent vacuum.
As visualization help, the reader can find animations for both processes in the links provided below \footnote{ Husimi Q-function time evolution under dynamics provided by the microscopic \url{http://www.hambrientosvagabundos.org/mpg/Micro.mp4} and the phenomenological \url{http://www.hambrientosvagabundos.org/mpg/Pheno.mp4} approaches.}.

We conducted an analysis for initial squeezed coherent states of the field but the dynamics are similar to those for coherent states, the mean value of the quadratures follow a spiral decay to the coherent vacuum and, at short times, the variances of the quadratures equalize and follow a behaviour equivalent to that of coherent states.

\section{Conclusions}

We have derived the microscopic master equation for the Jaynes-Cummings model under field dissipation at finite temperature and off-resonance. 
We revisited evolution in the well-known zero-$T$ single-excitation manifold, where the difference in the dynamics under the microscopic and phenomenological approaches appear as a high-frequency modulation of the ground state probability in the phenomenological approach, constructed an analytic closed form for the state evolution, and show the effect of detuning between the qubit and field frequencies on the effective decay rates; for initial states with an excited qubit a larger detuning produces a lower decay rate and the opposite for initial states with the qubit in the ground state.
This is obvious, due to the decay channels, and deliver a consequent ordering of the qubit-field purity minima. Interestingly enough, these minima are not  observed in the von Neumann entropy for the field or the concurrence of the joint qubit-field state that show a high-frequency modulation.

We confirmed numerically these behaviours beyond the single-excitation manifold at finite temperatures for initial Fock states of the field, where the dynamics start in a well-defined excitation manifold, and studied dissipation for initial coherent states, where the dynamics start in an extended superposition of excitation manifolds.
For initial coherent states of the field, dynamics under the microscopic approach provides a faster suppression of the collapse and revivals of the population inversion than the phenomenological approach, but the real difference is observed in phase-space, where the microscopical approach provides a smooth spiral decay trajectory of the field quadratures, while the phenomenological approach produces more convoluted dynamics with highly oscillating variances in the quadratures.

In summary, while a phenomenological treatment makes it simpler to create a building block approach to open systems that does not differ much at short times from the predictions of a formal treatment, a microscopic treatment of dissipation produces smoother dynamics that are closer to what semi-classical intuition might signal.
This seems to suggest that it becomes imperative to follow formal approaches to dissipation in order to describe multipartite interaction models. 

\vspace{6pt} 

\ack{	
	Carlos A. Gonz\'alez-Guti\'errez is grateful to CONACYT for financial support under doctoral fellowship No. 385108 and acknowledges the assistance of Ing. Israel Rebolledo in elaborating the ion-trap picture showed in Fig. \ref{fig:Fig1}.
	B. M. Rodr\'iguez-Lara thanks Alexander Moroz for fruitful discussion, acknowledges funding through CONACYT CB-2015-01-255230 grant, and thanks the Photonics and Mathematical Optics Group for hosting him. 
}

\section*{References}

\end{document}